\documentclass[aps,prd,preprint,tightenlines,superscriptaddress,nofootinbib]{revtex4}
\usepackage{epsfig}
\usepackage{graphicx}
\usepackage{psfrag}
\usepackage{amsmath,amssymb}
\usepackage{colordvi}
\usepackage{amsfonts}
\usepackage{enumerate}
\usepackage{slashed}
\usepackage{color}
\usepackage{xcolor}

\begin{document}

\title{Soft Factor Subtraction and Transverse Momentum Dependent Parton Distributions on Lattice}

\author{Xiangdong Ji}
\affiliation{INPAC, Department of Physics, and Shanghai Key Lab for Particle Physics and Cosmology,
Shanghai Jiao Tong University, Shanghai, 200240, P. R. China}
\affiliation{Center for High-Energy Physics, Peking University, Beijing, 100080, P. R. China}
\affiliation{Maryland Center for Fundamental Physics, University of Maryland, College Park, Maryland 20742, USA}
\author{Peng Sun}
\affiliation{Nuclear Science Division, Lawrence Berkeley
National Laboratory, Berkeley, CA 94720, USA}
\author{Xiaonu Xiong}
\affiliation{Center for High-Energy Physics, Peking University, Beijing, 100080, P. R. China}
\author{Feng Yuan}
\affiliation{Nuclear Science Division, Lawrence Berkeley
National Laboratory, Berkeley, CA 94720, USA}
\date{\today}
\vspace{0.5in}
\begin{abstract}
We study the transverse momentum dependent (TMD) parton distributions in 
the newly proposed quasi-parton distribution function framework in Euclidean
space. A soft factor subtraction is found to be essential to make the TMDs
calculable on lattice. We show that the quasi-TMDs with the associated soft
factor subtraction can be applied in hard QCD scattering 
processes such as Drell-Yan lepton pair production in hadronic collisions.
This allows future lattice calculations to provide information on
the non-perturbative inputs and energy evolutions for the TMDs.
Extension to the generalized parton distributions and quantum
phase space Wigner distributions will lead to a complete nucleon
tomography on lattice.
\end{abstract}

\maketitle

{\it Introduction.}
Transverse momentum dependent (TMD) parton distributions have attracted great attentions 
in hadron physics research~\cite{Boer:2011fh}. They provide a unique aspect
of the partonic structure of nucleon, by extending the conventional
description of Feynman parton distribution to the transverse dimension.
Experimental investigations of the TMDs in the last few years
have stimulated great theoretical developments, such as the QCD
factorization, the energy (scale) evolution, and the universality of the TMDs~\cite{Collins,Collins:1981uk,Ji:2004wu,bbdm}.
These developments have laid solid foundation in phenomenological applications
in hard QCD processes and hence allow us to extract the TMDs from the experiments. 

In addition to the experimental accesses, the lattice QCD shall also be able to compute the TMDs. 
This has become evidently possible, due to a recent breakthrough proposal
to calculate the parton distributions (PDFs) in Euclidean space in the large momentum
effective theory approach~\cite{Ji:2013dva}. 
To compute the parton distributions on lattice is conceptually 
important in the applications of QCD in hadron physics~\cite{Ma:2014jla,Xiong:2013bka,Lin:2014zya,Musch:2010ka}.
Progresses have been made concerning the technique issues associated with 
the applications of the quasi-PDFs, mainly on the integrated parton
distributions~\cite{Ma:2014jla,Xiong:2013bka,Lin:2014zya}, 
Early attempts of formulating TMDs on lattice
have been performed in Refs.~\cite{Musch:2010ka}, where the 
TMD definitions were modified and only the moments were considered. 
In this paper, we will apply the idea of Ref.~\cite{Ji:2013dva} to the TMD
case and, in particular, build the QCD factorization description of the hard
scattering process such as Drell-Yan lepton pair production in $pp$ collisions.
The goal is to identify the TMD operators which can be
computed on lattice and applied in QCD hard process 
in a consistent and rigorous fashion. 
A key point of the effective theory approach is that 
the theoretical uncertainties are under control~\cite{Ji:2013dva}.

In addition, the lattice calculations can also help the phenomenological applications of the TMDs.
In particular, if we want to make predictions for future experiments, not only the TMDs at lower scale
but also the relevant energy evolutions become important~\cite{Aybat:2011ta,Sun:2013dya}.
In previous phenomenological studies, various assumptions are 
made~\cite{Sun:2013dya,Aidala:2014hva,Aybat:2011zv,Landry:2002ix,Echevarria:2014xaa}
and they differ from each other. If we can compute the TMDs on lattice, it 
will provide important guidelines for the phenomenological studies.

In the following, we will carry out an explicit one-loop perturbative calculation for the TMDs, 
and demonstrate the QCD factorization in terms of the quasi-TMDs for the Drell-Yan process.
By doing so, we will find that a soft factor subtraction in the TMD definition is essential
to fulfill the factorization argument. The soft factor is constructed in such a way that it
can be computed on lattice. Therefore, the TMDs with the soft factor 
subtraction are desirable for the lattice computation in the future. 
This provides a solid foundation for future lattice applications for the
TMDs and many other distributions, such as the generalized parton distributions and 
quantum phase space Wigner distributions.

Soft factor is an important feature in the TMD factorization of hard QCD processes, 
which is also related to the regulation 
for the light-cone singularity in the TMD parton distributions. In the literature,
there have been several proposals, and each of them introduces a
way to construct the soft factor in the final factorization formula~\cite{Collins,Collins:1981uk,Ji:2004wu,bbdm}. 
Following the quasi-PDF framework, we will derive 
a unique soft factor subtraction.
Most importantly, both TMDs and the soft factor can be computed
on lattice.

The rest of this paper is organized as follows. In Sec.II, we introduce
the definition of the TMDs in Euclidean space, and will show the
soft factor subtraction is necessary. In Sec. III, we apply the TMDs
to the Drell-Yan process and show that the QCD factorization
at one-loop order can be achieved, where the soft factor plays an essential role. 
We briefly discuss the Collins-Soper
evolution of the TMDs in Sec. IV and conclude our paper in Sec. V.

{\it TMD Definition and Soft Factor Subtraction.}
For convenience, we consider the proton moving in $+\hat z$ direction with momentum,
\begin{equation}
P=\Lambda p+\frac{M^2}{2\Lambda}n \ ,
\end{equation}
where $\Lambda=P^+$ is a large momentum scale and $M$ is the proton mass, and we have introduced
two light-like vectors $p=(0^-,1^+,0_\perp)=\bar n$ and $n=(1^-,0^+,0_\perp)$:
$p^2=n^2=0$ and $p\cdot n=1$.
We further introduce a space-like vector $n_z=\frac{1}{\sqrt{2}}\left(n-p\right)$, such that
$n_z^2=-1$. $P_z$ is related to the projection of $P$ along with $n_z$,
$n_z\cdot P=-P_z $.
In the limit of $P_z\gg M$ or massless case, we have $\Lambda=\sqrt{2}P_z$.

In applying the TMD parton distributions and the associated QCD factorization, we keep the
leading power contribution in the limit of $P_z\gg k_\perp$ where $k_\perp$
is the transverse momentum. We neglect all higher power corrections of $k_\perp/P_z$.
This power counting analysis is consistent with the large momentum effective
theory arguments to compute the parton distribution on lattice~\cite{Ji:2013dva}.
In this framework, the TMD quark distribution is  written as, 
\begin{equation}
q(x_z,k_\perp)=\frac{1}{2}\int\frac{d^3z}{(2\pi)^3}e^{ik\cdot z}\langle PS|\overline{\psi}(0)
{\cal L}_{n_z(0,-\infty)}^\dagger\gamma^z{\cal L}_{n_z(z,-\infty)}\psi(z)|PS\rangle \ , \label{tmdq}
\end{equation}
where $x_z=k_z/P_z$. In the above definition, ${\cal L}_{n_z(y,-\infty)}={\cal P}exp\left\{-ig\int_0^{-\infty}
d\lambda n_z\cdot A(\lambda n_z+y)\right\}$ represents
the gauge link along the $\hat z$ direction~\footnote{We focus our discussions in
covariant gauge. In a singular gauge, such as the axial gauge $n_z\cdot A=0$, we have to include 
an extra gauge link in the spatial infinity~\cite{Ji:2002aa}.}. 
It has been known that the TMDs are process-dependent, and we have chosen
the gauge link path to $-\infty$ which indicates that the above definitions are for
the Drell-Yan process. 
In the TMD factorization, the cross section
and the parton distributions are conveniently written in the
so-called impact parameter space, which is the Fourier
transformation respect to the transverse momentum:
$q(x_z,b_\perp)=\int \frac{d^2k_\perp}{(2\pi)^2} e^{-ik_\perp\cdot b_\perp} q(x_z,k_\perp)$.

The TMD quark distribution defined as Eq.~(\ref{tmdq}) contains
the soft gluon radiation, which has to be subtracted in the final factorization
formula. Similar to the idea proposed by Collins in Ref.~\cite{Collins},
we introduce the following subtraction,
\begin{equation}
q^{sub.}(x,b_\perp)=q^{unsub.}(x,b_\perp)\sqrt{\frac{S^{n_x,n_y}(b_\perp)}{S^{n_x,n_z}(b_\perp)S^{n_z,n_y}(b_\perp)}} \ , \label{tmd}
\end{equation}
where $q^{unsub.}(x_z,b_\perp)$ is the un-subtracted PDF
in Eq.~(\ref{tmdq}) and $S$ is defined as
\begin{equation}
S^{\bar v, v}(b_\perp)={\langle 0|{\cal L}_{\bar v(-\infty,0)}^\dagger(b_\perp)
{\cal L}_{v(0,-\infty)}^\dagger(b_\perp) {\cal
L}_{ v(0,-\infty)}(0) {\cal L}_{\bar v(-\infty,0)}(0) |0\rangle   }\, , \label{softg}
\end{equation}
with ${\cal L}_v$ the gauge link to infinity along the direction $v$. 
In the above subtraction, we have chosen two transverse
Wilson lines: $n_x^2=n_y^2=-1$ and $n_x\cdot n_z=n_y\cdot n_z=n_x\cdot n_y=0$, 
to construct the associated soft factor. From the factorization point of view, light-like vectors for $n_{x,y}$ could be
used as well. However, such soft factor can not be calculated on lattice.

An important consequence of subtracting the soft factor from the TMDs
defined in Eq.~(\ref{tmdq}) is that the gauge link self-interaction diagrams 
(such as Figs.~1(c) and 2(c) shown in the following section) are canceled out by the similar
contribution from the soft factor of the last term in Eq.~(\ref{tmd}). These
diagrams, in general, can introduce a pinch singularity~\cite{bbdm}
in the TMD calculations, and the subtraction is essential to fulfill the
factorization for the associated hard processes~\footnote{
Similar idea can work out for the case of the TMD 
factorization studied in Ref.~\cite{bbdm} extending the original Collins-Soper
81 definition of the TMDs in axial gauge~\cite{Collins:1981uk} to a
covariant gauge.}.

After the subtraction, the TMD quark distribution of Eq.~(\ref{tmd}) is well 
defined and calculable on lattice. This kind of subtraction method
in lattice QCD has been applied earlier in the literature, see, for example,
Ref.~\cite{Creutz:1980wj}. This technique will have profound implications
in the quasi-PDFs framework. In the following calculations, we will first 
focus on the applications in the TMD factorization. 

\begin{figure}[tbp]
\begin{center}
\includegraphics[width=10cm]{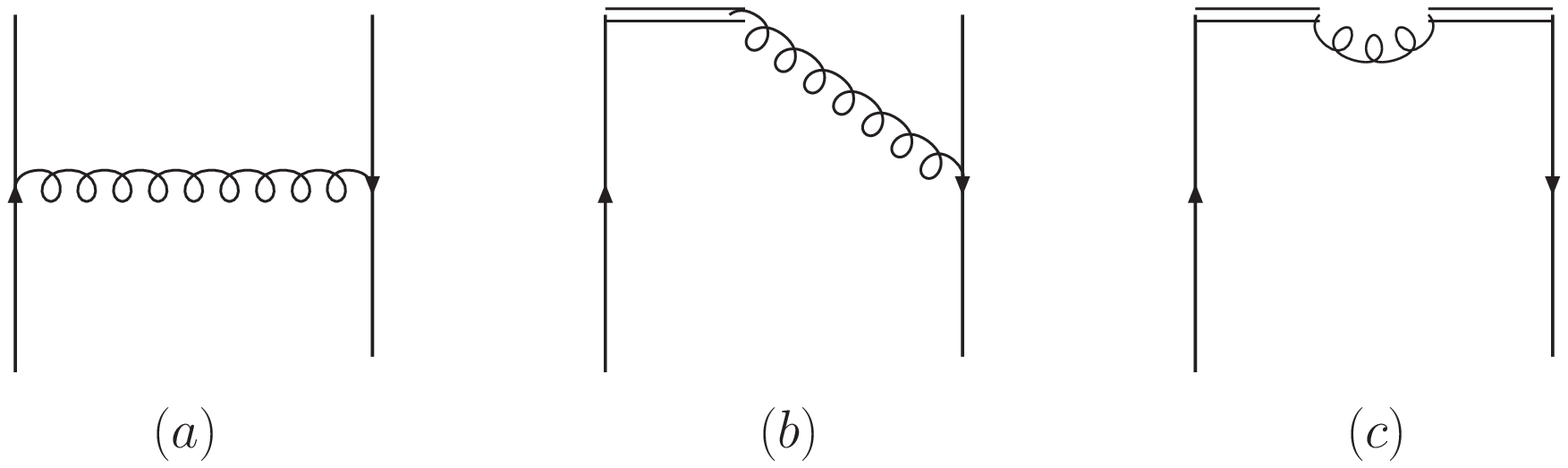}
\end{center}
\caption[*]{Real diagrams contributions to the TMD quark distributions at one-loop order.}
\label{pdfr}
\end{figure}

\begin{figure}[tbp]
\begin{center}
\includegraphics[width=10cm]{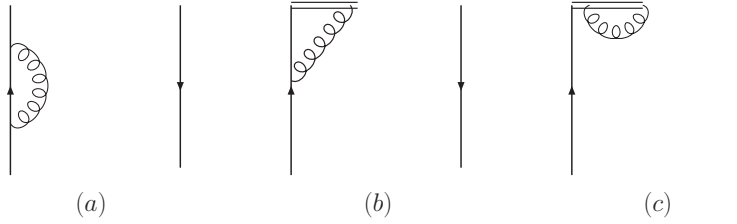}
\end{center}
\caption[*]{Virtual diagrams contributions to the TMD quark distributions at one-loop order.}
\label{pdfv}
\end{figure}

{\it One-loop Calculations and Factorization in Drell-Yan Process.}
It is instructive to have one-loop calculations and investigate the associated
factorizations in terms of the new TMDs defined in the last section. 
For the one-loop calculations, we take an on-shell quark target. Clearly, the leading order
quark distribution can be written as
$q^{(0)}(x_z,k_\perp)=\delta (1-x_z)\delta^{(2)}(k_\perp)$, which leads to the 
expression in the impact parameter space: $q^{(0)}(x_z,b_\perp)=\delta(1-x_z) $.
One-loop corrections contain real and virtual diagrams as shown in Figs.~(\ref{pdfr},\ref{pdfv}).
First, the calculations for the virtual diagrams are similar to those in Ref.~\cite{Ji:2004wu,Sun:2013dya}.
We only need to change $\zeta^2\to -\zeta^2$ where $\zeta^2=(2n_z\cdot P)^2/(-n_z^2)$.

For the real diagrams, different from the conventional TMDs, the quasi-TMDs
will have contributions from $x_z>1$ region, similar to that calculated for the 
integrated parton distributions~\cite{Ma:2014jla,Xiong:2013bka}. 
However, these contributions are power suppressed in the limit of $k_\perp\ll P_z$. 
For example, Fig.~\ref{pdfr}(a) contributes,
\begin{equation}
\frac{\alpha_s}{4\pi^2}C_F\frac{1-\epsilon}{k_\perp^2}\frac{(1-x_z)\left(\sqrt{k_\perp^2+P_z^2(1-x_z)^2}+P_z(1-x_z)\right)}{\sqrt{k_\perp^2+P_z^2(1-x_z)^2}} \ .
\end{equation}
Clearly, the contribution in the 
region of $x_z>1$ is power suppressed. Therefore, in the limit of $P_z\gg k_\perp$, it reduces to
\begin{equation}
\frac{\alpha_s}{2\pi^2}C_F\frac{1-\epsilon}{k_\perp^2}(1-x_z) \ ,
\end{equation}
for $0<x_z<1$, where $2\epsilon=4-D$ with $D$ the dimension. In our calculations, we take
dimension regulation to regulate singularities in both ultra-violet and
infra-red regions. 
Fig.~\ref{pdfr}(b) contributes,
\begin{equation}
\frac{\alpha_s}{4\pi^2}C_F\frac{1}{k_\perp^2}\frac{x_z}{1-x_z}
\frac{\left(\sqrt{k_\perp^2+P_z^2(1-x_z)^2}+P_z(1-x_z)\right)}{\sqrt{k_\perp^2+P_z^2(1-x_z)^2}} \ . \label{e6}
\end{equation}
Again, the contribution in the region of $x_z>1$ is also power suppressed. However, there is
a singularity at $x_z=1$. In order to evaluate the leading power contribution, we introduce a plus function and 
take the limit of $k_\perp^2\ll P_z^2$, 
\begin{equation}
\frac{\alpha_s}{2\pi^2}C_F\frac{1}{k_\perp^2}\left(\frac{2x_z}{(1-x_z)_+}+\delta(1-x_z)\ln\frac{\zeta^2}{k_\perp^2}\right) \ .\label{e7}
\end{equation}
To derive the above result, we have taken into account the fact that there are
contributions below and above $x_z=1$ in Eq.~(\ref{e6}), and a principal 
value prescription has been applied to evaluate the second term in Eq.~(\ref{e7}).
After this procedure, the leading power contributions are again limited
to the region of $0<x_z\le 1$.

As we discussed in the previous section, Figs.~$1(c )$ and $2(c )$ will
be cancelled out by similar diagrams from the soft factor subtraction, and they 
will not contribute to the subtracted TMDs. By adding all contributions, we
find that the total TMD quark distribution at one-loop order, will be
\begin{eqnarray}
q^{sub.}(x_z,b_\perp;\zeta)&=&\frac{\alpha_s}{2\pi}C_F\left\{\left(-\frac{1}{\epsilon}+\ln\frac{c_0^2}{b_\perp^2\bar\mu^2}\right){\cal P}_{q\to q}(x_z)+(1-x_z)\right.\nonumber\\
&&\left.+\delta(1-x_z)\left[\frac{3}{2}\ln\frac{b_\perp^2\mu^2}{c_0^2}+\ln\frac{\zeta^2}{\mu^2}
-\frac{1}{2}\left(\ln\frac{\zeta^2b_\perp^2}{c_0^2}\right)^2-2\right]\right\} \ , \label{oneloop}
\end{eqnarray}
in the impact parameter $b_\perp$-space, where $c_0=2e^{-\gamma_E}$ and ${\cal P}_{q/q}(x)$ is the usual splitting kernel for the quark.

To apply the above TMD quark distribution in hard QCD process
as Drell-Yan lepton pair production in $pp$ collisions, we need to calculate the TMD antiquark 
distribution as well. Similar to $\zeta$ introduced
above, for the antiquark distribution we introduce 
the energy parameter $\bar\zeta^2=(2n_z\cdot \bar P)^2/(-n_z^2)$
where $\bar P$ is the momentum for the hadron moving in the $-\hat 
z$ direction. The differential cross section depending on the transverse momentum
of the lepton pair can be written in the following factorization form,
\begin{equation}
W(Q,b_\perp)=q^{sub.}(x_z,b_\perp;\zeta)\overline{q}^{sub.}(\bar x_z,b_\perp;\bar\zeta)H(Q,\mu) \ ,
\end{equation}
in the impact parameter space. The Fourier transform of the above $W(Q,b_\perp)$ will
lead to the transverse momentum distribution of the differential cross section. 
From the factorization,
we obtain the hard factor for the Drell-Yan process as
\begin{equation}
H(Q)=\frac{\alpha_s}{2\pi}C_F\left[\ln\frac{Q^2}{\mu^2}+\pi^2-4\right] \ ,
\end{equation}
where we have chosen $\zeta^2=\bar\zeta^2=Q^2$ for simplicity~\footnote{After resummation
of large logarithms by solving the Collins-Soper evolution 
equations, the dependence on $\zeta$ and $\bar\zeta$ will
cancel out and lead to the unique final results for the 
differential cross sections depending on the 
transverse momentum.}. 

{\it TMD evolution.} Similar to the previous formalisms for the TMDs, the TMDs in the quasi-parton
distribution framework in Euclidean space also depend on the energy of the hadron. 
This can be seen from the one-loop calculations in the last section, in particular, 
from a double logarithms term $\ln^2(\zeta^2b_\perp^2)$
as shown in Eq.~(\ref{oneloop}). 
Following previous examples, we can derive the Collins-Soper evolution
equation for the TMDs in the quasi-parton distribution framework too,
\begin{equation}
\frac{\partial}{\partial \ln\zeta} q(x_z,b_\perp,\zeta)=\left(K(b_\perp,\mu)+G(\zeta,\mu)\right)
\times q(x_z,b_\perp,\zeta) \ , \label{cs}
\end{equation}
where $K$ and $G$ are the soft and hard parts, respectively. 
Except a power counting analysis~\cite{Korchemsky:1994is},
we do not know much about the soft part in the evolution kernel.
This has caused the model dependences in the phenomenological
studies in the literature~\cite{Sun:2013dya,Aidala:2014hva,Aybat:2011zv,Landry:2002ix,Echevarria:2014xaa}.

If we can compute the TMDs from lattice QCD, we will be
able to extract the evolution information directly from the lattice. 
In practice, we need to perform the lattice calculations for several different
values of $P_z$, and we can calculate numerically the dependence of the TMDs
on the energy of the hadrons. 
This is of importance for phenomenological
applications, for example, to investigate the energy dependence of the 
Sivers asymmetries, which is one of the top questions in hadronic spin physics.

{\it Conclusion and Discussions.}
In summary, we have shown that the proposed framework of Ref.~\cite{Ji:2013dva}
for parton distributions can be applied to the transverse momentum dependent 
parton distributions, where the soft factor subtraction plays a very important
role. We have calculated the TMDs at one-loop order, and
demonstrated the associated factorization for the Drell-Yan lepton 
pair production. We further showed that the
Collins-Soper evolution can be derived for the TMDs in Euclidean space,
and argued that the future lattice simulation will be important for the theory predictions
for hadronic observables.

We would like to emphasize that the soft factor subtraction is crucial to
achieve the factorization. More importantly, this soft factor can be calculated
from lattice. Future lattice QCD calculations of the TMDs can serve as important
inputs for hard processes, and can also be used to study the parton distribution
in three-dimension fashion. Extending to the quantum phase space Wigner distributions
is straightforward, which, in return, will provide computational access to the nucleon
tomography in parton picture. 

In the above calculations, we have shown the perturbative calculations at one-loop
order. It will be interesting to extend to higher order corrections. From the generic factorization
argument, we expect the soft factor subtraction will be also important 
to understand the quasi-PDFs at two-loop and beyond. We will carry 
out the detailed analysis in a future publication.

In addition, the soft factor subtraction deals with the self-interaction 
diagrams from the gauge links in the quasi-PDF definition as those
in Fig.~$1(c )$ and Fig.~$2(c )$. The existence
of these diagrams come from the fact that the gauge links are along the
non-light-like directions. The contributions of these diagrams lead to
subtle ultra-violet behaviors in the perturbative calculations at one-loop order,
which have to be carefully handled in the matching between the quasi-PDFs and
the conventional ones~\cite{Xiong:2013bka,Lin:2014zya,Ma:2014jla}.
Since the subtraction method introduced above is very general, and should apply to 
various parton distributions.
This will help the convergence of the matching calculations in
these papers.

We thank J.~Collins, M.~Engelhardt, J.~W.~Qiu for discussions and comments.
This work was partially supported by the U.
S. Department of Energy via grants DE-FG02-93ER-40762
and DE-AC02-05CH11231 and a grant from National Science Foundation of China (X.J.).


\begin{thebibliography}{99}

\bibitem{Boer:2011fh}
  D.~Boer 
  {\it et al.},
  arXiv:1108.1713 [nucl-th];
  A.~Accardi
  {\it et al.},
  arXiv:1212.1701 [nucl-ex].
  
  
  \bibitem{Collins}
J.C.Collins, {\it Foundations of Perturbative QCD}, Cambridge University Press, Cambridge, 2011.

\bibitem{Ji:2004wu}
  X.~Ji, J.~P.~Ma and F.~Yuan,
  Phys.\ Rev.\ D {\bf 71}, 034005 (2005).

\bibitem{Collins:1981uk}
J.~C.~Collins and D.~E.~Soper,
Nucl.\ Phys.\ B {\bf 193}, 381 (1981) [Erratum-ibid.\ B {\bf 213},
545 (1983)];
Nucl.\ Phys.\ B {\bf 197}, 446 (1982).

\bibitem{bbdm} A.~Bacchetta, D.~Boer, M.~Diehl and P.~J.~Mulders,
  JHEP {\bf 0808}, 023 (2008).

\bibitem{Ji:2013dva}
  X.~Ji,
Phys.\ Rev.\ Lett.\  {\bf 110}, 262002 (2013);
  arXiv:1404.6680 [hep-ph].


\bibitem{Ma:2014jla} 
  Y.~-Q.~Ma and J.~-W.~Qiu,
  arXiv:1404.6860 [hep-ph].



\bibitem{Xiong:2013bka} 
  X.~Xiong, X.~Ji, J.~-H.~Zhang and Y.~Zhao,
  arXiv:1310.7471 [hep-ph].
  
\bibitem{Lin:2014zya} 
  H.~-W.~Lin, J.~-W.~Chen, S.~D.~Cohen and X.~Ji,
  arXiv:1402.1462 [hep-ph].

  
\bibitem{Musch:2010ka}
  B.~U.~Musch, P.~Hagler, J.~W.~Negele and A.~Schafer,
  Phys.\ Rev.\ D {\bf 83}, 094507 (2011);
  B.~U.~Musch, P.~.Hagler, M.~Engelhardt, J.~W.~Negele and A.~Schafer,
  Phys.\ Rev.\ D {\bf 85}, 094510 (2012).




  
\bibitem{Aybat:2011ta} 
  S.~M.~Aybat, A.~Prokudin and T.~C.~Rogers,
  Phys.\ Rev.\ Lett.\  {\bf 108}, 242003 (2012).
  
\bibitem{Sun:2013dya} 
  P.~Sun and F.~Yuan,
  Phys.\ Rev.\ D {\bf 88}, 034016 (2013);
  Phys.\ Rev.\ D {\bf 88}, 114012 (2013).
  
\bibitem{Aidala:2014hva} 
  C.~A.~Aidala, B.~Field, L.~P.~Gamberg and T.~C.~Rogers,
  Phys.\ Rev.\ D {\bf 89}, 094002 (2014).

\bibitem{Echevarria:2014xaa} 
  M.~G.~Echevarria, A.~Idilbi, Z.~-B.~Kang and I.~Vitev,
  Phys.\ Rev.\ D {\bf 89}, 074013 (2014).
  


\bibitem{Aybat:2011zv}
  S.~M.~Aybat and T.~C.~Rogers,
  Phys.\ Rev.\ D {\bf 83}, 114042 (2011).


\bibitem{Landry:2002ix}
  F.~Landry, R.~Brock, P.~M.~Nadolsky and C.~P.~Yuan,
  Phys.\ Rev.\ D {\bf 67}, 073016 (2003);
  Phys.\ Rev.\ D {\bf 63}, 013004 (2001).
  
  


\bibitem{Ji:2002aa} 
  X.~Ji and F.~Yuan,
  Phys.\ Lett.\ B {\bf 543}, 66 (2002);
  A.~V.~Belitsky, X.~Ji and F.~Yuan,
  Nucl.\ Phys.\ B {\bf 656}, 165 (2003).
  
\bibitem{Creutz:1980wj} 
  M.~Creutz,
  Phys.\ Rev.\ Lett.\  {\bf 45}, 313 (1980).
\bibitem{Korchemsky:1994is} 
  G.~P.~Korchemsky and G.~F.~Sterman,
  Nucl.\ Phys.\ B {\bf 437}, 415 (1995).
  
  
  
\end{thebibliography}
\end{document}